\documentstyle[aps]{revtex}

\tightenlines

\begin{document}
\title{Theory of vortex matter transformations in the high $T_{c}$
superconductor YBCO }
\author{Dingping Li$^{\text{1,2}}$ and Baruch Rosenstein$^{\text{2}}$}
\address{$^{\text{1}}${\it National Laboratory of Solid State
Microstructures and Department of Physics}\\
{\it Nanjing University, Nanjing, 210093,China}\\
$^{\text{2}}${\it Electrophysics Department and National Center for
Theoretical Sciences}\\
{\it National Chiao Tung University, Hsinchu 30050, Taiwan, R. O.
C.}}
\date{\today }
\maketitle

\begin{abstract}
Flux line lattice in type II superconductors undergoes a transition{\bf \ }%
into a \textquotedblleft disordered\textquotedblright\ phase like vortex
liquid or vortex glass, due to thermal fluctuations and random quenched
disorder. We quantitatively describe the competition between the thermal
fluctuations and the disorder using the Ginzburg -- Landau approach. The
following $T-H$ phase diagram of $YBCO$ emerges. There are just two distinct
thermodynamical phases, the homogeneous and the crystalline one, separated
by a single first order transitions line. The line however makes a wiggle
near the experimentally claimed critical point at $12T$. The
\textquotedblleft critical point\textquotedblright\ is reinterpreted as a
(noncritical) Kauzmann point in which the latent heat vanishes and the line
is parallel to the $T$ axis. The magnetization, the entropy and the specific
heat discontinuities at melting compare well with experiments.
\end{abstract}




The $H-T$ phase diagram of a high $T_{c}$ superconductor is very complex due
to the competition between thermal fluctuations (TF) and disorder. At low
fields vortex solid melts into a liquid \cite{Blatter} due to TF. The
discontinuity in magnetization \cite{Zeldov,Willemin,Nish} and in the
specific heat \cite{Schilling,Bouquet} unambiguously demonstrate that the
transition is a first order. Evidence is growing that the solid-glass
transition in YBCO \cite{Kokkaliaris} is also a first order\cite%
{Nish,Radzyner}. It was suggested \cite{Nish} that the two first order phase
transition lines $H_{\ast }(T)$ and $H_{m}(T)$ join at a multicritical point
or that the melting line continues as a second order transition between the
putative liquid II and a liquid\cite{Bouquet}. However other experiments are
interpreted using a concept of ``unified first order transition line'' \cite%
{Ertas}: only character of the transition evolves from the thermal
fluctuation dominated to the disorder dominated one without multicritical
points along the line. This single line was demonstrated in anisotropic
materials {\it BSCCO} \cite{Avraham}, {\it LaSrCuO} \cite{Radzyner2} and
{\it NdCeCuO} \cite{Radzyner} and was claimed recently in {\it YBCO}\cite%
{Pal}.

Theoretically idealized models like the frustrated $XY$ model \cite%
{Olsson,HuMC} or a collection of interacting pointlike objects subject to
both the pinning potential and the thermal bath Langevin force were
simulated numerically\cite{Otterlo,Reichhardt}. Alternatively the melting
line was located using the phenomenological Lindemann criterion on the solid
side either in the framework of the ``cage''\ model \cite{Ertas} or a more
sophisticated approach based on the elasticity theory\cite%
{Kierfeld,Giamarchi}. Although both approaches are very useful in more
``fluctuating''\ superconductors like {\it BSCCO}, a problem arises with
their application to {\it YBCO} close $T_{c}$: vortices are far from being
pointlike and even their cores significantly overlap. As a consequence the
behavior of a dense vortex matter is different from that of a system of
pointlike vortices and of the $XY$ model.

In this Letter we quantitatively study the effects of the competition of
thermal fluctuations and disorder in the framework of the anisotropic
Ginzburg -- Landau (GL) model appropriate precisely in the region of
interest for $YBCO$. It was used successfully to describe the ''clean'' case%
\cite{Thouless,Li}.

The model without disorder is defined by free energy\cite{Blatter}:

\begin{equation}
\int {d^{3}}x\frac{\hbar ^{2}}{2m_{ab}}\left| {\left( {\nabla -\frac{2\pi i}{%
\Phi _{0}}A}\right) \psi }\right| ^{2}+\frac{\hbar ^{2}}{2m_{c}}\left| {%
\partial _{z}\psi }\right| ^{2}-\alpha T_{c}(1-t)\left| \psi \right| ^{2}+%
\frac{\beta }{2}\left| \psi \right| ^{4},  \label{GL}
\end{equation}%
where $\Phi _{0}=hc/(2e)$, $t=T/T_{c}$, ${\bf A}=(By,0,0)$. The model
provides a good description of thermal fluctuations as long as $1-t-b<<1$,
where $b=H/H_{c2}$. An ``applicability line''\ $1-t-b<<0.2$ \ is depicted on
Fig.1. The model however is highly nontrivial even without disorder and
within the lowest Landau level (LLL) approximation \cite{Thouless} in which
only the LLL mode is retained and the free energy simplifies (after
rescaling) :
\begin{equation}
g_{LLL}=\int d^{3}x\left[ \frac{1}{2}|\partial _{z}\psi |^{2}+a_{T}|\psi
|^{2}+\frac{1}{2}|\psi |^{4}\right] .  \label{GLscal}
\end{equation}%
The simplified model has just one parameter -- the (dimensionless) scaled
temperature: $a_{T}\equiv (t+b-1)\sqrt[3]{2/(Gi\pi ^{2}b^{2}t^{2})}$ with
the Ginzburg number $Gi\equiv 32\left( {\pi \lambda ^{2}T_{c}\gamma /(\Phi
_{0}^{2}\xi )}\right) ^{2}$, ${\gamma }^{2}{=}m_{c}/m_{ab}$ the anisotropy
parameter, ${\lambda }$ the magnetic penetration depth and $\xi $ the
coherence length. The (effective) LLL model is applicable in a surprisingly
wide range of fields and temperatures determined by the condition that the
relevant excitation energy $\varepsilon $ is much smaller than the gap
between Landau levels $\varepsilon _{c}\equiv 2\hbar eB/(cm_{ab})$. Within
the mean field approximation in the liquid $\varepsilon /\varepsilon _{c}=x$
is a solution of the ``gap equation''\cite{Lawrie}): $2bx+\left(
1-t-b\right) =2\pi t\sqrt{Gib}(\sqrt{4x-2}+\sum_{n=0}^{\infty }(\sqrt{4x+4n+2%
}-\sqrt{4x+4n-2}-1/\sqrt{n+x})].$ The LLL dominance line in Fig.1 represents
a conservative condition $\varepsilon /\varepsilon _{c}=1/20$. On the right
of the line, the HLL (higher Landau level) TF modes should be considered,
while for $b<(1-t)/13$, HLL in solid becomes important\cite{Li}. Apart from
the fields smaller than $H_{LLL}\approx 3kG$, experimentally observed the
melting and the solid-glass lines are well within the range of applicability
of the LLL approximation.

The disorder is described within the GL approach via a random field $W_{x}$
with correlator $\overline{\text{\ }W_{x}W_{y}}=W\delta (x-y)$\cite{Blatter}%
, while TF are described by partition function $Z=\int_{\psi ^{\ast },\psi
}\exp \left\{ -\left[ F+\int_{x}W_{x}\left| {\psi }_{x}\right| ^{2}\right]
/T\right\} $. Assuming small $W$, the free energy $-T\ln Z$ is calculated
perturbatively to the second order:
\begin{equation}
g=g_{cl}+\int_{x}W_{x}\left\{ \left\langle \left| {\psi }_{x}\right|
^{2}\right\rangle -\;\frac{1}{2T}\int_{y}W_{y}\left[ \left\langle \left| {%
\psi }_{x}\right| ^{2}\left| {\psi }_{y}\right| ^{2}\right\rangle
-\left\langle \left| {\psi }_{x}\right| ^{2}\right\rangle \left\langle
\left| {\psi }_{y}\right| ^{2}\right\rangle \right] \right\} ,
\end{equation}%
where $\left\langle {}\right\rangle $ and $g_{cl}$ denote the thermal
average and effective free energy of \ the clean system. Averaging random
field one obtains in the scaled units $g=g_{cl}-n(t)\Delta g_{dis}$, where $%
\Delta g_{dis}=\frac{1}{2}\int_{x}\left[ {\left\langle {\left| {\psi }%
_{x}\right| ^{4}}\right\rangle -\left\langle {\left| {\psi }_{x}\right| ^{2}}%
\right\rangle }^{2}\right] $, and $n(t)\propto W/T$.

Recently a quantitative theory \cite{Li} of TF in ''clean'' vortex liquids
and solids based on the GL model of eq.(\ref{GLscal}) was developed and was
successfully applied to the fully oxidized $YBa_{2}Cu_{3}O_{7}$ for which
the effects of disorder are negligible in whole currently accessible fields
range\cite{Blatter,Nish}. The two loop result for solid in the clean case $%
g_{sol}=-a_{T}^{2}/(2\beta _{A})+2.85\left| {a_{T}}\right| ^{1/2}+2.4/a_{T}$
is sufficiently precise (on the 0.1\% level) all the way to melting. We
calculated the one loop disorder correction: $\Delta g_{sol}=2.14\left| {%
a_{T}}\right| ^{1/2}.$ An explicit expression for $g_{liq}(a_{T})$, obtained
using the Borel-Pade resummation of the renormalized high temperature series
is rather bulky and can be found in \cite{Li}. The disorder correction in
liquid can be obtained by differentiating the ''clean'' partition function
with respect to parameters: $\Delta g_{liq}=(g_{liq}-2a_{T}g_{liq}^{\prime
})/3-(g_{liq}^{\prime })^{2}/2$. These results enable us to find the
location of the transition line and calculate discontinuities of various
physical quantities across the transition line.

{\bf {\it \ }}It was noted \cite{Li} that in a clean system a homogeneous
state exists as a metastable overcooled liquid state all the way down to
zero temperature. This is of importance since interaction with disorder can
convert the metastable state into a stable one. Indeed generally a
homogeneous state gains more than a crystalline state from pinning, since it
can easier adjust itself to the topography of the pinning centers. Since at
large $\left| {a_{T}}\right| $, $\Delta g_{liq}\propto a_{T}^{2}$ is larger
than $\Delta g_{sol}\propto \left| {a_{T}}\right| ^{1/2},$ the transition
line shifts to lower fields. The equation for the melting line is $%
d(a_{T})\equiv (g_{liq}-g_{sol})/(\Delta g_{liq}-\Delta g_{sol})=n(t).$ The
universal function $d(a_{T})$, plotted in Fig.2 turns out to be
non-monotonic. Since $n(t)$ is a monotonic function of $t$, one obtains the
transition lines for various $n$ in Fig.1 by ``sweeping''\ the Fig.2. A
peculiar feature of $d(a_{T})$ is that it has a local minimum at $%
a_{T}\approx -17.2$ and a local maximum at $a_{T}\approx -12.1$ (before
crossing zero at $\ a_{T}\approx -9.5)$. Therefore between these two points
there are three solutions to the melting line equation. As a result,
starting from the zero field at $T_{c},$ the transition field $H(T)$ reaches
a maximum at $E_{ent}$ (point {\it CP }in Fig{\it .1}) beyond which the
curve sharply turns down (this feature was called ``inverse melting''\ in %
\cite{Avraham}) and at $E_{mag}$ backwards. Then it reaches a minimum and
continues as the solid (Bragg glass) -- vortex glass line roughly parallel
to the $T$ axis.

The temperature dependence of the disorder strength $n(t)$, as of any
parameter in the GL approach, should be derived from a microscopic theory
assuming random chemical potential or fitted to experiment. We conjecture
(and find to be consistent with the experiments) that the general dependence
near $T_{c}$ is: $n(t)=n(1-t)^{2}/t$ (or $W\propto (1-t)^{2}$). The
expression approaches the one used lower temperatures \cite{Blatter} with $%
n=\gamma \gamma _{T}^{0}/(4\pi ^{2}\sqrt{2Gi}\xi ^{3})$. The best fit for
the low field part of the experimental melting line $H_{m}(T)$ of the
optimally doped {\it YBCO} (data taken from \cite{Bouquet}, $T_{c}$ $=92.6$,
$\ \gamma =8.3$) gives $Gi=2.010^{-4}$, $H_{c2}=190\ \ T$, \ $\kappa
=\lambda /\xi =50$ (consistent with other experiments\cite{Nish,Deligiannis}%
). This part is essentially independent of disorder. The upper part of the
melting curve is very sensitive to disorder: both the length of the
``finger''\ and its slope depend on $n$. The best fit is $n=0.12.$ This
value is of the same order of magnitude as the one obtained
phenomenologically using eq.(3.82) in \cite{Blatter}. We speculate that the
low temperature part of the ``unified''\ line corresponds to the solid --
vortex glass transition $H^{\ast }(T)$ observed in numerous experiments \cite%
{Nish,Bouquet,Kokkaliaris,Radzyner,Pal} (squares in Fig.1\cite{Nish}). A
complicated shape of the ``wiggling''\ line has been recently observed\cite%
{Pal}.

Magnetization and specific heat of both solid and liquid can be calculated
from the above expressions for free energy. Magnetization of liquid along
the melting line $H_{m}(T)$ is larger than that of solid. The magnetization
jump is compared in Fig 3a with the SQUID experiments \cite{Schilling}\ in
the range $80-90K$ (triangles) and of the torque experiments (stars \cite%
{Willemin} and circles \cite{Nish}). One observes that the results of the
torque experiments compare surprisingly well above $83K$ while the SQUID
data a bit lower than theoretical or the torque one. But those of \cite{Nish}
vanish abruptly below $83K$ unlike the theory and are inconsistent with the
specific heat experiments discussed below\cite{Bouquet,Deligiannis}. We
predict that at lower temperatures (somewhat beyond the range investigated
experimentally so far) magnetization reaches its maximum and changes sign at
the point $E_{mag}$ (at which magnetization of liquid and solid are equal).

Entropy jump calculated using the Clausius -- Clapeyron relation is compared
with an experimental one deduced from the spike of the specific heat \cite%
{Bouquet} (triangles in Fig. 3b) and an indirect measurement from the
magnetization jumps \cite{Nish} (circles). At high temperatures the
theoretical values are a bit lower than the experimental and both seem to
approach a constant at $T_{c}$. The theoretical entropy jump and the
experimental one of \cite{Bouquet} vanish at $E_{ent}$ (Fig.1) near $75K$.
Experimentally such a point (called Kauzmann points\cite{Debenedetti}) was
established in {\it BSCCO} as a point at which the ``inverse melting''\
appears\cite{Avraham}. Below this temperature entropy of the liquid becomes
smaller than that of the solid. Note that the equal magnetization point $%
E_{mag}$ is located at a slightly lower field than the equal entropy point $%
E_{ent}$. The Kauzmann point observed at a lower temperature in {\it YBCO}
in \cite{Radzyner} is different from $E_{ent}$ since it is a minimum rather
than a maximum of magnetic field. It is also located slightly outside the
region of applicability of our solution. The point $E_{ent}$ is observed in %
\cite{Pal} in which the universal line is continuous.

In addition to the spike, the specific heat jump has also been observed
along the melting line $H_{m}(T)$\cite{Schilling,Bouquet,Deligiannis}.
Theoretically the jump does not vanish either at $E_{ent}$ or $E_{mag}$, but
is rather flat in a wide temperature range. Our results are larger than
experimental jumps of \cite{Bouquet} by a factor of 1.4 to 2 (Fig.3c). In
many experiments there appears a segment of the second order phase
transition continuing the first order melting line beyond a certain point.
In \cite{Bouquet} it was shown that at that point the specific heat profile
shows ''rounding''. We calculated the specific heat profile above the
universal first order transition line. It exhibits a ''rounding''\ feature
similar to that displayed in \cite{Bouquet} with no sign of the criticality.
The height of the peak is roughly of the size of the specific heat jump. We
therefore propose not to interpret this feature as an evidence for a second
order transition above the first order line.

As seen from the fitting the optimal doping YBCO, the disorder parameter $n$
in the cases of interest is very small. However to address a more delicate
question of appearance of the glass transition between liquid and glass
claimed in some experiments \cite{Nish,Bouquet} and simulations \cite%
{Otterlo,HuMC}, we used a nonperturbative method, the replica trick,
combined with the gaussian variational approximation\cite{Mezard}. Breaking
of the replica symmetry signals appearance of the glass transition in static
phenomena. Here we summarize the results, leaving details for a longer
presentation. The most general hierarchical homogeneous (liquid) Ansatz \cite%
{Mezard} was considered and found that there is no replica symmetry breaking
(RSB) solution.

Now we compare our results with other theories starting with models based on
the Lindemann criterion \cite{Ertas,Kierfeld,Giamarchi}. Near $T_{c}$ (where
effects of disorder are small) the location of the first order transition
line $H(T)$ is qualitatively consistent with that found from the Lindemann
criterion \cite{Blatter}. In the intermediate region around $E_{ent}$ our
results for the melting line are completely different from all the
phenomenological models \cite{Ertas,Kierfeld,Giamarchi}. In particular we do
not have a critical point assumed in some of them (the two papers in \cite%
{Kierfeld} differ on the continuation of the line beyond the critical point
in which two first order lines join together). At temperatures below $70K$
the elasticity theory approach based on the London approximation \cite%
{Kierfeld,Giamarchi} (valid beyond the range of applicability of the GL
approach) is expected to smoothly interpolate to the GL approach. The
advantage of the present approach is that, in addition to location of the
transition line, it enables the determination of discontinuities and the
calculation of physical quantities away from the transition line.

The comparison with numerical simulations can be made only qualitatively
since the disordered GL model has not been simulated. In the XY simulations
of \cite{Olsson} a single transition line is parallel to the $T$ axis below
certain temperature, while ours is not. In \cite{HuMC} there is, in
addition, a slush - liquid transition, while Langevin simulation \cite%
{Otterlo} finds the liquid - glass transition. Their transition criterion
however is dynamical. Absence of RSB in liquid does not generally imply that
this state, especially at low temperatures, does not exhibit ``glassy
properties''\ in dynamics. In fact overcooled liquids generally are
``glassy''. Therefore we propose to consider the glassy properties of the
vortex state above the $H^{\ast }(T)$ line in the context of the
``disordered overcooled''\ liquids. Recent simulations \cite{Reichhardt}
demonstrate that glassy features in dynamics do not necessarily correspond
to the conventional vortex glass scenario. Next we comment on the ''vortex
loops''\ scenario for $YBCO$. The region in which the loops are relevant
according to \cite{Tesanovic} is below $GiH_{c2}$, too small to explain the
experimental transitions (our fit for $Gi=2$ $10^{-4}$ is much smaller than
that assumed in \cite{Tesanovic}). The LLL is inapplicable in this region
(see Fig.1). In strongly anisotropic materials, a \ model of the Lawrence --
Doniah type is more appropriate \cite{Blatter,Hu}. We performed a 2D GL
calculation and found that in that case there is no ``wiggle''\ of the
transition line and speculate that it disappears at certain value of the
anisotropy parameter.

We are grateful to E.H. Brandt, X. Hu and Y. Yeshurun for discussions, T.
Nishizaki, F. Bouquet, G. Mikitik, E. Zeldov and Z. Tesanovic for
correspondence. Supported by NSC of Taiwan grant NSC\#91-2112-M-009-503 and
the Mininstrey of Science and Technology of China (G1999064602) are
acknowledged.

\begin{figure}[tbp]
\caption{ Theoretical first order transition lines for various
degrees of disorder separated between a homogeneous and a
crystalline phases of vortex matter. The best fit $n=0.12$ line is
compared with experimental melting line $H_m (T)$
(\protect\cite{Bouquet}) and the solid - vortex glass transition
line$H^\ast (T)$ (\protect\cite{Nish}). Equal entropy and
magnetization points are denoted by $E_{ent} $and $E_{mag} $. The
range of
applicability is restricted by the two lines. Very small fields below $%
H_{LLL} $ are beyond this range. }
\end{figure}

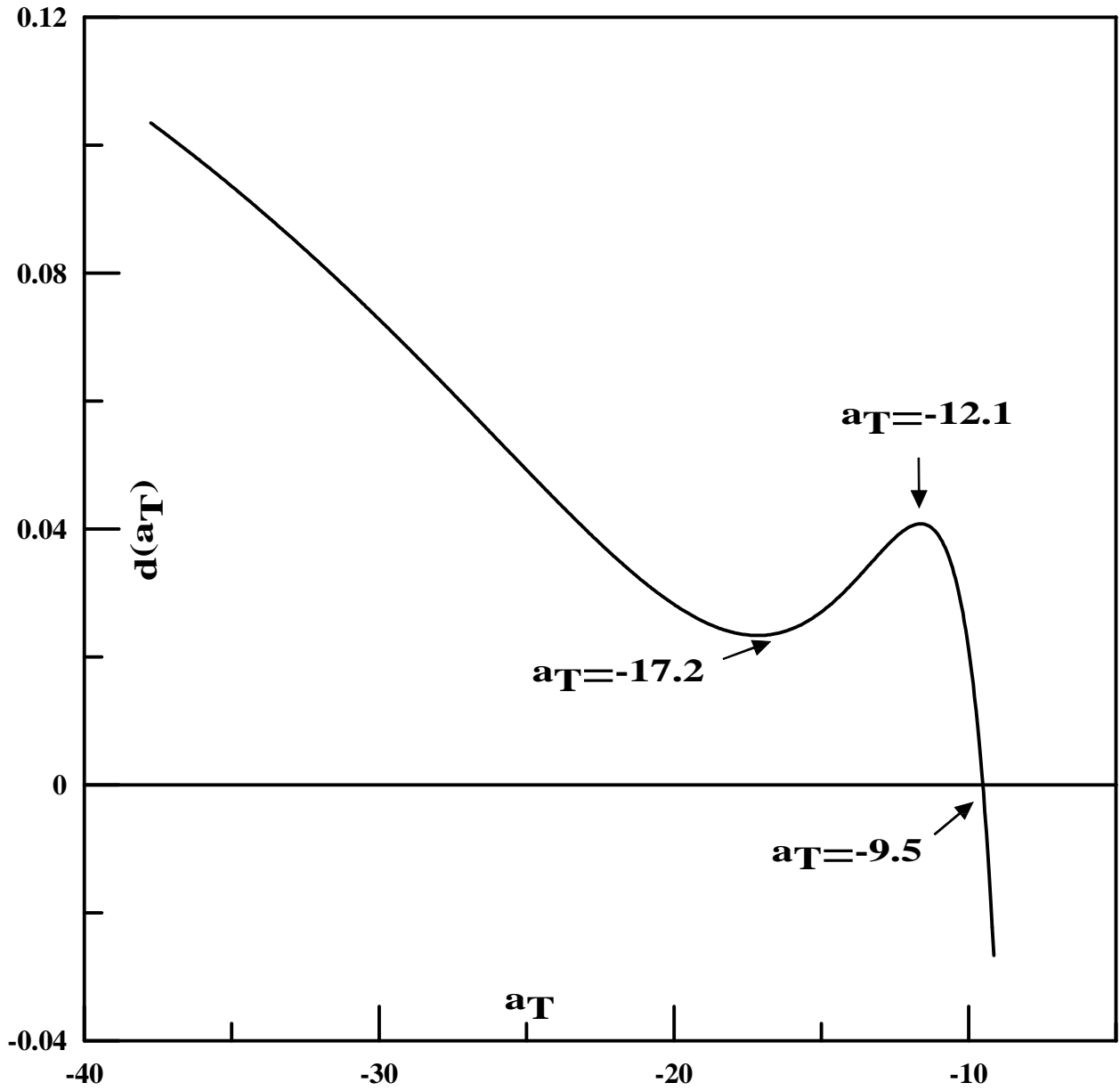
\begin{figure}[tbp]
\caption{ Non-monotonic dependence of the disorder function $d(a_{T})$ on
the LLL scaled temperature $a_{T}$}
\end{figure}

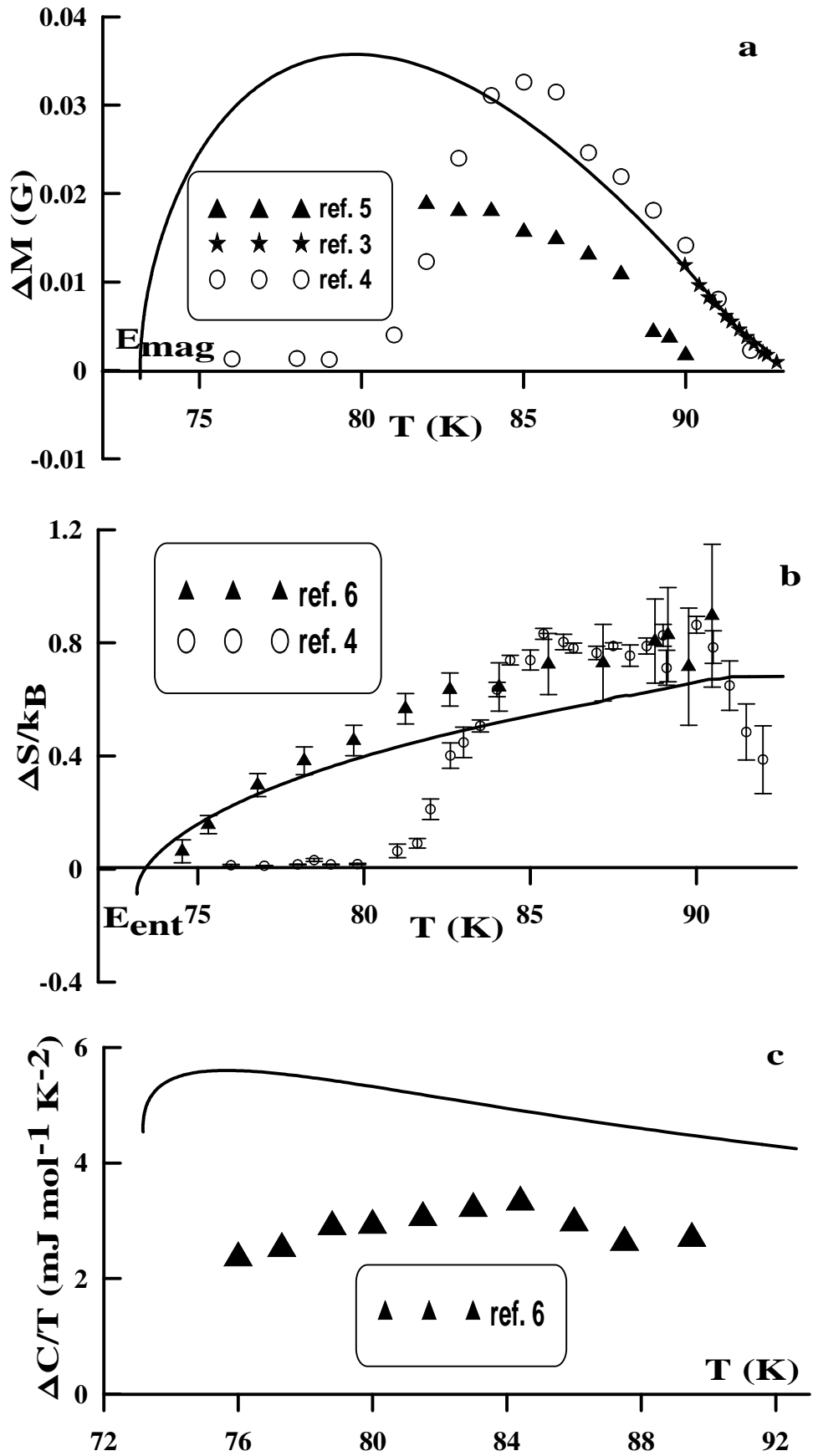
\begin{figure}[tbp]
\caption{ Discontinuities along the first order transition line in {\it YBCO}%
. Jumps of magnetization, entropy and specific heat are compared with
experiments in {\bf a}, {\bf b}, {\bf c} respectively. }
\end{figure}

\end{document}